\documentclass[14pt]{article}

\usepackage{amsmath,amsfonts,amsthm,amsbsy}
\usepackage{graphicx}
\usepackage{color}

\numberwithin{equation}{section}
\def\cite#1{[\ref{#1}]}

\def\a{\alpha}

\def\zb{\overline{z}}
\def\zb1{\overline{z}_{1}}
\def\zb2{\overline{z}_{2}}
\def\z1{z_{1}}
\def\z2{z_{2}}

\def\d{\delta}

\def\g{\gamma}

\def\l{\lambda}

\def\b{\beta}
\def\ha{\frac 12}
\def\cL{\cal L}

\theoremstyle{definition}

\begin{document}

\date{\today}

\vspace{0.5in}

\setcounter{page}{0}

{\bf \Large \center{Comparison of methods to determine point-to-point resistance in nearly rectangular networks with application to a ``hammock" network.}}

\hspace {0.5 in}

{\center{John W. Essam$^1$, Nikolay Sh. Izmailyan$^{2,3}$, Ralph Kenna$^2$ and Zhi-Zhong Tan$^4$}}

\hspace {0.5 in}

{\flushleft{
$^1$ Department of Mathematics, Royal Holloway College, University of London, Egham, Surrey TW20~0EX, England.

\hspace {0.5 in}

$^2$ Applied Mathematics Research Centre, Coventry University, Coventry CV1~5FB, England.

\hspace {0.5 in}

$^3$ Yerevan Physics Institute, Alikhanian Brothers 2, 375036 Yerevan, Armenia.

\hspace {0.5 in}

$^4$ Department of Physics, Nantong University, Nantong 226019, China.

\hspace {0.5 in}}}

\hspace {0.5 in}

{\bf \Large \center{Abstract}}

\hspace {0.5 in}

Considerable progress has recently been made in the development of  techniques to exactly determine two-point resistances in networks of various topologies. 
In particular, two types of  method have emerged. 
One is based on potentials and the evaluation of eigenvalues and eigenvectors of the Laplacian matrix associated with the network or its minors.
The second method is based on a recurrence relation associated with the distribution of currents in the network.
Here, these methods are compared and used to determine the resistance distances between any two nodes of a network with topology of a hammock.

\newpage
\section{Introduction}

The computation of {\emph{two-point resistances}} in  networks is a classical problem in electric circuit theory and graph theory, with applications in the study of transport in disordered media \cite{disorder}, random walks \cite{DS}, first-passage processes \cite{Redner} and lattice Green’s functions \cite{Kat}.
In recent decades, and especially in recent years, the problem has received widespread interest in the  mathematical, physical, engineering and chemical sciences because of its relevance to such a broad range of problems. 
A nice interpretation of the two-point resistance $R_{ij}$ between nodes $i$ and $j$ in a graph was given by Klein and Randic \cite{KR} as a novel distance function,  sometimes called the {\emph{resistance distance}} between nodes $i$ and $j$.
The term was used because of the associated physical interpretation: for unit resistors on each edge of a graph, $R_{ij}$ is small when there are many paths between the nodes i and j, and large when there are few paths between the nodes i and j. 
The total effective resistance, also called the {\emph{Kirchhoff index}} \cite{KR}, \cite{K}, was introduced in chemistry as an improved alternative to other parameters used for distinguishing different molecules with similar shapes. 
Also, it has been shown that the first passage time of random walks (the expected time to pass a special node for a walker starting from a source node) is related to the {\emph{effective resistance}} \cite{DS}. 
However, it is usually very difficult to obtain resistance distances in large complex graphs.

Different methods have recently been developed to compute the two-point resistances or resistance distances.
These include using the eigenvalues and eigenvectors of the Laplacian matrix  associated with the network \cite{XG},\cite{W},  the eigenvalues and eigenvectors of the minors of the Laplacian matrix \cite{IKW},\cite{IK}, the determinants of submatrices of the Laplacian matrix \cite{SR},\cite{SS},\cite{BGX}, and  recursion relations \cite{T11},\cite{TZY},\cite{TEW},\cite{YK}.

\begin{figure}[t]
\begin{center}
\includegraphics[scale=0.7]{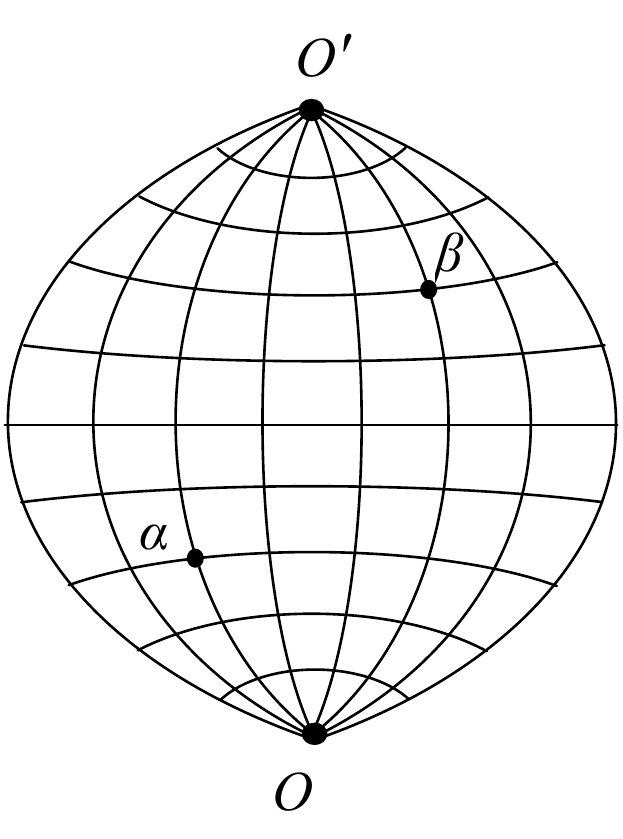}
\caption{\it A ``hammock'' with $M=9$ and $N=8$}   \label{hammock}
\end{center}
\end{figure}

Here we  compare two recently developed methods \cite{IK} and \cite{T11},\cite{TEW} by focusing on the determination of the point-to-point resistance in a rectangular resistor network with two additional nodes. 
Each of these nodes (denoted $O$ and $O^\prime$ in the hammock network of figure \ref{hammock}) is connected to all of the nodes in one of two opposing edge rows .
With free boundary conditions at the two edge columns, we refer to the network as a ``hammock". 
Nodes along the $M$ rows are connected by resistors of strength $r$ while nodes along the $N$ columns are linked by resistance $s$. 
The additional resistors connecting to $O$ and $O^\prime$ also have resistance $s$.

The first method presented here is based on evaluating eigenvalues and eigenvectors of the Laplacian matrix \cite{W} or first order \cite{IKW} or second order \cite{IK} minors of the Laplacian matrix associated with the resistor network. 
In what follows we will refer to this method as the {\bf \emph{Laplacian approach}} or {\emph{Method A}}. 
The second method \cite{TEW} is based on the solution of a recurrence relation  obtained by a matrix transformation of the equations relating the column currents. 
We refer to this method as the {\bf {\emph {Recursion-Transform (R-T) method}}} or {\emph{Method B}}.

The paper is organized as follows. 
In Sec.~\ref{descr} we briefly describe the two methods. 
In Sec.~\ref{results} we present a summary of our results for the point-to-point resistance for the ``hammock'' network.
We show how these results are obtained  by {\emph{Methods A}} and {\emph{B}} in Sec.~\ref{dervn}. 
We compare the two methods in Sec.~\ref{end}.

\section{Description of the two methods}
\label{descr}

\subsection{Method A: Using Laplacians}

If $J_i(k)$  is the current injected (or removed if negative) into the node at the intersection of row $i$ and column $k$, and if $V_k(i)$ is the corresponding potential, then, at interior nodes, current conservation gives 
\begin{equation}\label{Veq}
r^{-1} (V_{k+1}(i) -2 V_k(i)  + V_{k-1}(i) ) +s^{-1} ( V_k(i+1) -2V_k(i) + V_k(i-1)) = -J_i(k).
\end{equation}
Similarly, if  $J_0$ is the current at the added node $O$ and if $J_{M+1}$ is the current at added node $O^\prime$, and if these have potentials $V_0$ and $V_{M+1}$, respectively,
\begin{equation}
s^{-1}\sum_{k=0}^{N-1} (V_0 - V_k(1))=J_0 \qquad \hbox{and}\qquad s^{-1}\sum_{k=0}^{N-1} (V_{M+1}-V_k(M))=J_{M+1}.
\end{equation}
These equations are a special case of the formula for a resistor network consisting of $T$ nodes and  resistance $r_{i,j}=r_{j,i}$ between nodes $i$ and $j$, thus
\begin{equation}\label{simeq}
{\bf L}_T \bar V_T = \bar J_T,
\end{equation}
where $\bar V_T= \{V_0,V_1,\dots. V_{T-1}\}, \bar J_T= \{J_0,J_1,\dots J_{T-1}\}$ and the Laplacian matrix ${\bf L}_T$ has general element
\begin{equation}\label{LT}
{\bf L}_T (i,j)  = \left \{\begin{array}{lll}
c_{i,j} = -r_{i,j}^{-1}=c_{j,i}& \,\,\hbox{for}\,\, & i\ne  j\\
\\
c_{i,i}=\sum_{j=0}^{T-1} r_{i,j}^{-1}.
\end{array}
\right .
\end{equation}
The resistance between nodes $\alpha$ and $\beta$ ($R_{\alpha,\beta}$) can be written as \cite{W}
\begin{equation}
R_{\alpha,\beta}=\sum_{i=1}^{T-1}\frac{\left|\psi_{i\alpha}-\psi_{i\beta}\right|^2}{\lambda_i},
\label{Rab}
\end{equation}
where the $\lambda_i$ are the non-zero eigenvalues of  ${\bf L}_T$ and $\Psi_{i}=(\psi_{i,1},\psi_{i,2},...,\psi_{i,T-1})$ are the corresponding orthonormal eigenvectors.

The two-dimensional Laplacian of an $M\times N$ rectangular network with free boundaries can be written in terms of two one-dimensional Laplacians in the form
\begin{eqnarray}\label{decomp}
{\bf L}_{M \times N}^{{\rm{free}}}&=&r^{-1}{\bf L}_N^{{\rm{free}}}\otimes {\bf U}_M+s^{-1}{\bf U}_N \otimes {\bf L}_M^{{\rm{free}}}
\end{eqnarray}
where ${\bf L}_M^{{\rm{free}}}$ is the Laplacian for an $M$-node chain with free boundaries and $U_M$ is a unit matrix of dimension $M$.
 The Laplacian for other combinations of boundary conditions can be similarly written. The eigenvalues of the two-dimensional lattice matrix are therefore sums of the eigenvalues of the one-dimensional chain matrices and the eigenvectors are products of the corresponding eigenvectors.

The ``hammock'' Laplacian cannot be decomposed in this way, so that the eigenvalues and eigenvectors are far more difficult to determine. A similar problem arises if only one additional vertex is added to the rectangle, giving rise to a``fan" network \cite{IKfan}. However for the ``fan'' only $MN$ of the $MN+1$ equations \eqref{simeq} are independent   so setting the potential of the extra vertex to zero eliminates the corresponding row and column and  the Laplacian may be replaced by the resulting minor  which may be decomposed in the form \eqref{decomp}. The resulting equations are independent and all of the eigenvalues are non-zero and  are to be included in the sum \eqref{Rab}. 
This method is due to Izmailian, Kenna and Wu  who used it to determine the ``cobweb'' \cite{IKW} and ``fan'' \cite{IKfan} resistance.

If the free boundary conditions of the ``hammock'' are replaced by periodic ones, the resulting topology is known as a ``globe". 
The same problem arises in that the Laplacian cannot be decomposed. 
Izmailian and Kenna \cite{IK} showed that a solution was to replace the Laplacian by the minor $\cal L$ obtained by deleting the rows and columns corresponding to both added nodes. 
However a rather complicated correction needs to be made. The resulting formula for the resistance between any two nodes $\alpha$ and $\beta$ other than the node $0$ is
\begin{equation}
R_{\alpha,\beta}={\cal L}^{-1}_{\alpha,\alpha} + {\cal L}^{-1}_{\beta,\beta}  - {\cal L}^{-1}_{\alpha,\beta}- {\cal L}^{-1}_{\beta \alpha}+ \frac{\left(\sum_{i=1}^{MN}({\cal L}_{i,\alpha}^{-1}-{\cal L}_{i,\beta}^{-1})c_{0,i} )\right)^2}{c_0-\sum_{i=1}^{MN}\sum_{j=1}^{MN}{\cal L}_{ij}^{-1}c_{j,0}c_{0,i}}.
\label{resistorab}
\end{equation}
This expression is evaluated for the ``hammock'' in section \ref{sec:ham}.

\subsection{Method B}

\begin{figure}[t]
\begin{center}
\includegraphics[scale=1.0]{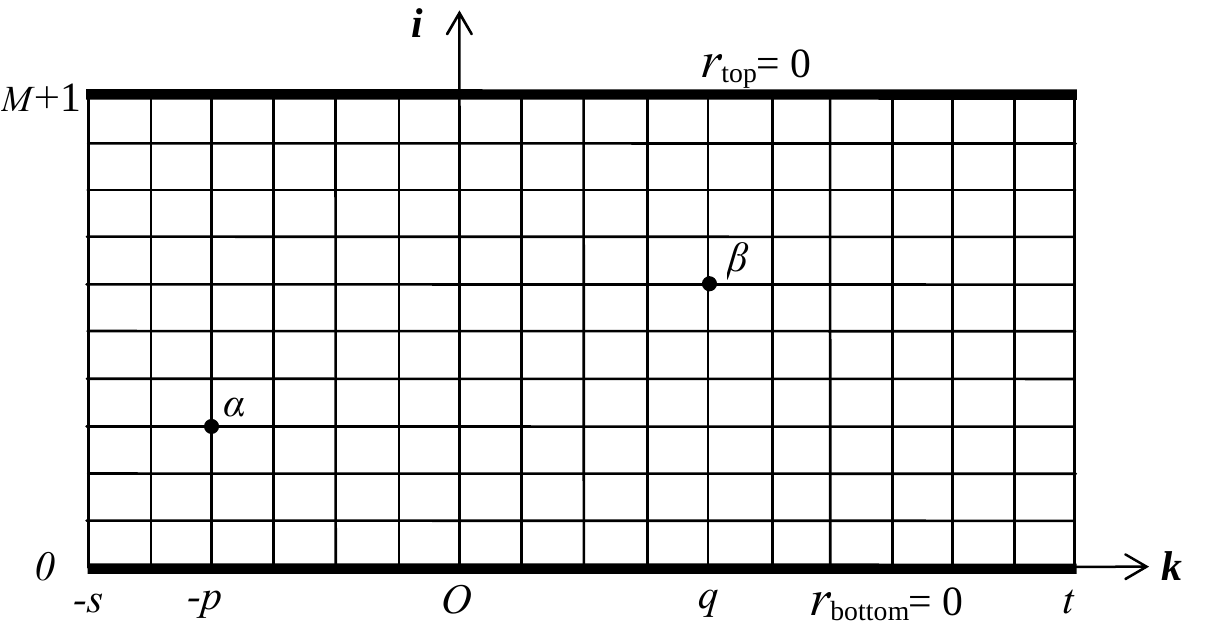}
\caption{{\it Rectangular network with the top and bottom rows of resistors having zero resistance.
 \newline $s=6, p=4, q=4, t=10, x_1=3, x_2=11, y_1=3, y_2=6, M=9, N=17$
\newline In method B, the input node $\a \equiv N_p$ and the output node  $\b \equiv N_q$ }}   
\label{hammock0}
\end{center}
\end{figure}

{\emph{Method B}} was introduced by Tan \cite{T11}. See also Tan, Zhou and Yang \cite{TZY}. 
Let $I_k(i)$ be the upward current in column $k$ between nodes at heights $i-1$ and $i$. 
We consider the ``hammock'' to be a rectangle with $N$ columns and $M+2$ rows with zero resistance in the top and bottom rows. 
The rows are labelled $i=0$ to $M+1$ (see figure \ref{hammock0}).

Suppose current $J$ is injected in column $k=z$ at height $i=y$.
In section \ref{sec:B:ham} we derive the following relation between the currents in three adjacent columns:
\begin{align}\label{Igen}
s I_{k+1}(i) =&-r_{i-1} I_k(i-1)+ (r_{i}+r_{i-1}+2s)I_k(i)   -r_{i} I_k(i+1) -sI_{k-1}(i)\nonumber\\
 &+J(r_{i}\d_{i,y}-r_{i-1}\d_{i,y+1})\d_{k,z}
\end{align}
where the resistors in row $i$ have resistance $r_i$. For the ``hammock'' $r_0=r_{M+1}=0$ and $r_i=r$ for $1\le i \le  M$.

With the definition $I_k= \{I_k(1),I_k(2),\dots,I_k(M+1)\}^T$ the current terms involving the radial resistance on the right of equation \eqref{Igen}  may be expressed in matrix form as $r{\bf L}^{{\rm{free}}}_{M+1}I_k$ where 
${\bf L}^{{\rm{free}}}_{M+1}\equiv 2 U_{M+1}-W_{M+1}$ is the  Laplacian matrix for a linear chain of
 $M+1$ nodes with free boundaries.  Here   $W_{M+1}$ is given by equation \eqref{Wm}.

The eigenvalues and eigenvectors of ${\bf L}^{{\rm{free}}}_m$ are known \cite{W}. Next we define a matrix $\Psi$, the rows of which are the eigenvectors of ${\bf L}^{{\rm{free}}}_{M+1}$, and use it obtain transformed current vectors $X_k \equiv \Psi I_k, k=1,2,\dots, N$. Applying $\Psi$ to the matrix form \eqref{Ieq} of equation \eqref{Igen} shows that for each row $i$,  $X_k(i)$ satisfies a separate second order recurrence on the column index $k$ (see \eqref{Xeq}). This may be solved in the standard way with two parameters in each of the three regions of $k$ delineated by the boundaries and the input and output nodes. 
Having determined the parameters by imposing the boundary conditions, the currents are obtained from the resulting $X_k$ using the inverse transformation $I_k =\Psi^{-1} X_k$. 
The potential difference, and hence the resistance, between the input and output nodes is obtained by summing the potential differences (determined by the currents) along a path between the nodes via the common node $i=M+1$ (see equation \eqref{Res}).


\section{Results for the resistance of the $M\times N$ ``hammock'' network  between two arbitrary nodes}\label{results}

We next present some new results for the ``hammock'' network coming from each approach. 
Then we present details of the derivations using each method. 
In the final section we compare the advantages and disadvantages of each approach.


\qquad
\subsection{ Notation}

\qquad{\bf Method A:} The nodes of the rectangular part of the ``hammock'' are labelled $(x,y)$ where $x = 1,2\dots N$ and $y=1,2\dots M$. The input and output nodes are $\a=(x_1,y_1)$ and $\b=(x_2,y_2)$ respectively. 
\qquad

{\bf Method B:} 
The ``hammock'' will be supposed to have $N=s+t+1$ radial lines, labelled from $k=-s$ to $k=t$.  
The input node $N_p \equiv \a$ is distant $y_1$ up the radial line $k=-p$ and the output node $N_q\equiv \b$ is distant $y_2$ up the radial line $k=q$.  Thus $x_1= s-p+1=N-t-p$ and $x_2 = s+q+1=N+q-t$. (see figure \ref{hammock0}).

\subsection{Main Result}  
 
Let ${\displaystyle{
u_i =2 + 2 h \left[{1-\cos\left (\frac{(i-1)\pi}{M+1} \right )}\right]}}$  and let $\l_i$ be the greater solution of \begin{equation}\label{laeq}
\l^2 -u_i \l+1=0
\end{equation}
where $h=r/s$ the horizontal-to-vertical resistance ratio.
 With
\begin{equation}\label{Lch}
L_i =\frac 12\ln \l_i  \qquad \hbox{or} \qquad \cosh(2L_i) = \frac 12 u_i,
\end{equation}
the resistance of the ``hammock'' between $N_p$ and $N_q$ is found to be
\begin{equation}
R_{x_1,y_1}^{x_2,y_2}= \frac{2r}{M+1}\sum_{i=2}^{M+1}\frac {\a  S_i(y_1)^2 -2\beta  S_i(y_2)S_i(y_1) +\gamma  S_i(y_2)^2}{\sinh(2L_i)\sinh(2N L_i)} +\frac{s(y_2-y_1)^2}{N(M+1)}
\label{Rhammockasym}
\end{equation}
where ${\displaystyle{S_i(y) = \sin{\left[{\frac{(i-1)\pi y}{M+1}}\right]}}}$.
Here,  in the notation of method A
\begin{align}
\alpha  &=\cosh(2N-2x_1+1)L_i \cosh(2x_1-1)L_i,\nonumber\\
\beta  &=\cosh(2N-2x_2+1)L_i \cosh(2x_1-1)L_i,\nonumber\\
\gamma  &=\cosh(2N-2x_2+1)L_i \cosh(2x_2-1)L_i,\label{xylab}\\
\intertext{while in the notation of method B}
\alpha  &=\cosh(2t+2p+1)L_i \cosh(2s-2p+1)L_i,\nonumber\\
\beta  &=\cosh(2t-2q+1)L_i \cosh(2s-2p+1)L_i,\nonumber\\
\gamma  &=\cosh(2t-2q+1)L_i \cosh(2s+2q+1)L_i.\label{pqlab}
\end{align}

\subsection{Resistance between two nodes on the same radial line}

Without loss of generality we take the line to be $k=0$ and set $p=q=0$ so that $x_1=x_2=x$ say. In this case $\alpha = \beta = \gamma$ with the result
\begin{align}\label{RhammockRadial}
R_{x,y_1}^{x,y_2}=\frac{2r}{M+1} \sum_{i=2}^{M+1}{ [S_i(y_2)-S_i(y_1)]^2 \frac{\cosh[(2s+1)L_i]\cosh[(2t+1)L_i]}{\sinh[2L_i]\sinh[(2N L_i]}}\nonumber\\
+\frac{s(y_2-y_1)^2}{N(M+1)}.
\end{align}
Note that  $2s+1= 2x_1-1$ and $2t+1=2N-2x_2+1$.

\subsection{Resistance between two nodes on the same transverse line.}

Setting $y_1=y_2=y$, so that $S_i(y_1) = S_i(y_2) = S_i(y)$, the numerator of the summand in \eqref{Rhammockasym} becomes $(\a-2\b+\g)S_i(y)^2$. 
Further, if we set $p=q$ the distance between the input and output nodes is $d=2q = x_2-x_1$.
In this case
\begin{equation}
\a-2\b+\g = 2\sinh(dL_i)(\sinh [(2N-d)L_i] +\sinh(d L_i) \cosh [2(t-s)L_i]).
\end{equation}
If, furthermore, $s=t$, then the input and output nodes are symmetrically placed relative to the radial line boundaries and
\begin{equation}
R_{x_1,y}^{x_2,y}=\frac{4r}{M+1}\sum_{i=2}^{M+1}\frac { \sinh(d L_i)\cosh[(N-d)L_i]}{\sinh(2L_i)\cosh(NL_i)} S_i(y)^2.
\label{Rhammockasymtra}
\end{equation}
Eqs.(\ref{Rhammockasym}), (\ref{RhammockRadial}) and (\ref{Rhammockasymtra}) comprise the main new results of this paper. 
We next present their derivation using the two methods.
This will facilitate a comparison between the two approaches in Section~5.

\vspace{0.5 in}

\section{Derivation of the general form \eqref{Rhammockasym} by two different methods}
\label{dervn}

\subsection{Method A: Using the Laplacian approach \cite{IK}}\label{sec:ham}

\begin{figure}[h]
\begin{center}
\includegraphics[scale=0.8]{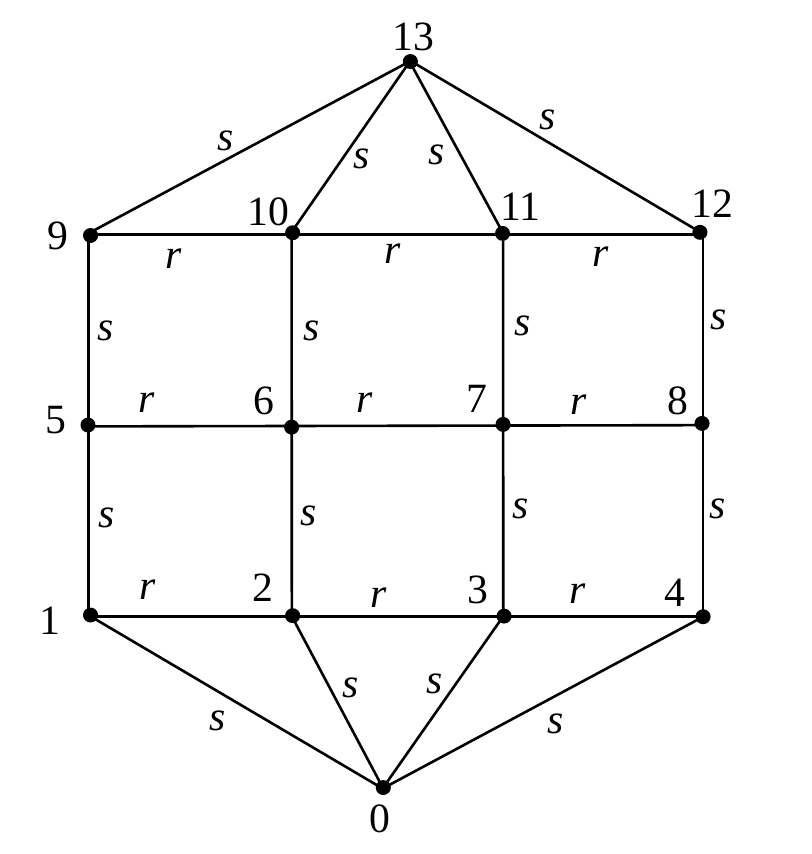}
\caption{\it A ``hammock'' network with $M=3$ and $N=4$ illustrating the coordinate labelling. }   
\label{34hammock}
\end{center}
\end{figure}

We begin with the expression \eqref{resistorab}  for the point-to-point resistance in terms of the Laplacian ${\cal L}_{ij}$ of the rectangular part of the "hammock". 
The nodes on the rectangular part are labelled by  $\{(x,y), x =1,2,\dots N, y=1,2,\dots,M\}$   so that in \eqref{resistorab} $i=x+(y-1)N$ (see figure~\ref{34hammock}).
Node~$(1,1)$ is positioned at the lower left hand corner.
 For the ``hammock'' network the elements of the first row and  column of the Laplacian \eqref{LT} have the following values
\begin{eqnarray}
c_{0,i}=c_{i,0}&=&s^{-1} \qquad \mbox{for} \quad i=1,2,3,...,N, \nonumber\\
c_{0,i}=c_{i,0}&=&0 \qquad \quad \mbox{for} \quad i=N+1,N+2,N+3,...,MN, \nonumber
\end{eqnarray}
and $c_0$ is given by
\begin{equation}
c_0=\sum_{j=1}^{MN+1}c_{0,j}=N s^{-1}.
\end{equation}
Equation \eqref{resistorab} can be transformed to
\begin{align}
R_{\alpha,\beta}&=
\frac{\Sigma_2^2}{N s - \Sigma_1}+ \hat R_{\alpha,\beta},\label{resistorab1}\\
\hbox{where} \hspace{0.3 in} \Sigma_1&=\sum_{i=1}^{N}\sum_{j=1}^{N}{\cal L}_{ij}^{-1},\hspace{0.3 in} \Sigma_2=\sum_{i=1}^{N}\left({\cal L}_{i,\alpha}^{-1}-{\cal L}_{i,\beta}^{-1}\right)\label{Sigmas}\\
\hbox{and}\hspace{0.3 in} \hat R_{\alpha,\beta}& ={\cal L}^{-1}_{\alpha,\alpha} + {\cal L}^{-1}_{\beta,\beta}  - {\cal L}^{-1}_{\alpha,\beta}- {\cal L}^{-1}_{\beta \alpha}\label{Rhat}.
\end{align}
${\cal L}^{-1}_{i,j}$  is the $(i,j)$th element of the inverse matrix ${\bf {\cal L}^{-1}}$ which  is given by
\begin{equation}\label{Lexpand}
{\cal L}^{-1}_{i,j}=\sum_{k=1}^{MN}\frac{\psi_{k,i}\psi_{k,j}^{*}}{\Lambda_k},
\end{equation}
where $\Lambda_k$ and $\psi_{k,i}$ are eigenvalues and eigenvectors of the second minor ${\bf {\cal L}}$ of the Laplacian.

\begin{figure}[t]
\begin{center}
\includegraphics[scale=1.0]{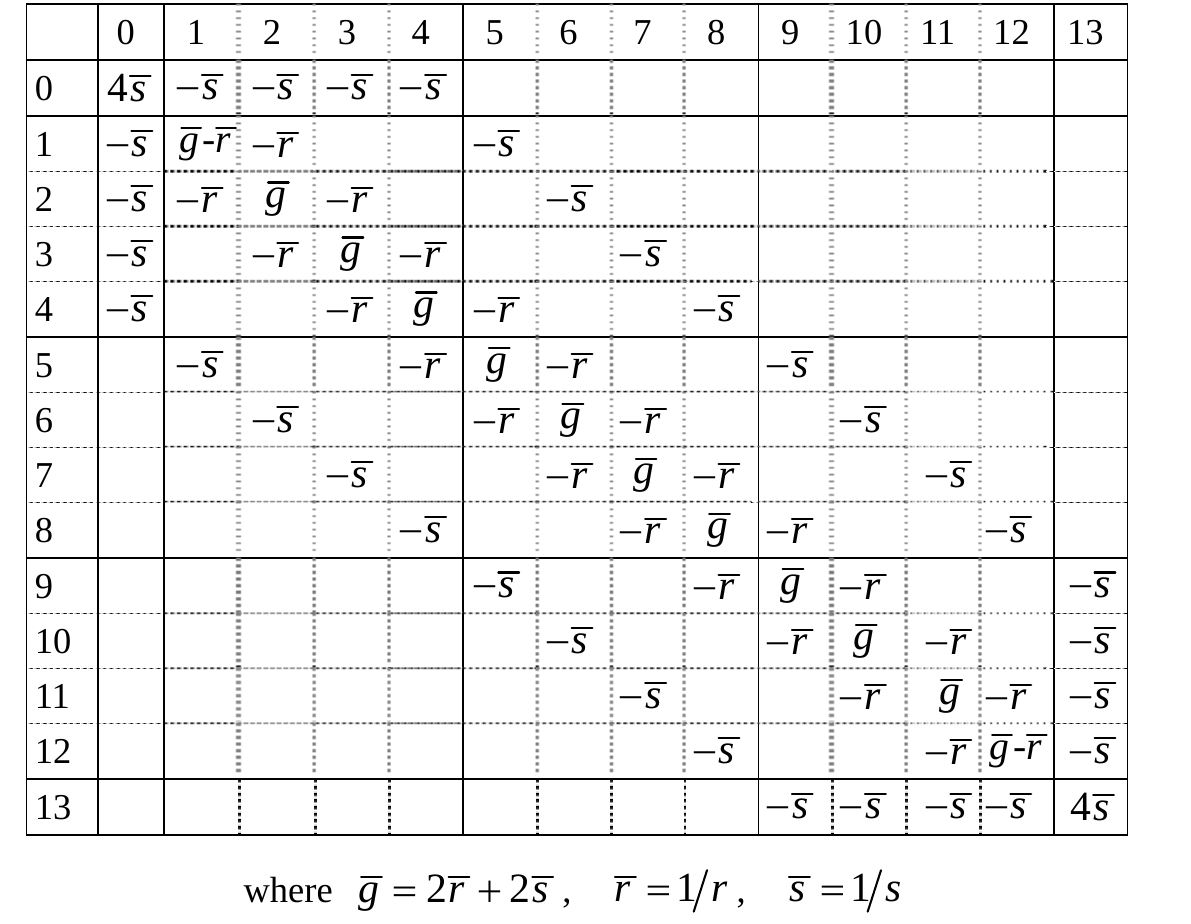}
\end{center}
\caption{The ``hammock'' Laplacian ${\bf L}_{3,4}$}
\label{laplace34}
\end{figure}
The second minor of the Laplacian for the ``hammock'' network may be
factored in a similar way to the rectangular network. For the example of figure~\ref{laplace34}
\begin{eqnarray}
{\bf {\cal L}}_{3 \times 4}^{\rm{hammock}} &=&
 s^{-1}\left[\begin{array}{rrr}
2&-1&0\\
-1&2&-1\\
0&-1&2
\end{array}\right]
\otimes \left[\begin{array}{rrrr}
1&0&0&0\\
0&1&0&0\\
0&0&1&0\\
0&0&0&1
\end{array}\right]
\nonumber \\
& +& r^{-1}\left[\begin{array}{rrrr}
1&0&0\\
0&1&0\\
0&0&1
\end{array}\right] \otimes
\left[\begin{array}{rrrr}
1&-1&0&0\\
-1&2&-1&0\\
0&-1&2&-1\\
0&0&-1&1
\end{array}\right]
\end{eqnarray}
or, in general,
\begin{equation}
{\bf {\cal L}}_{M \times N}^{\rm{hammock}}=s^{-1} {\bf
L}_M^{^{\rm{DD}}}\otimes{\bf U}_N +r^{-1} {\bf U}_M \otimes {\bf
L}_N^{\rm{{\rm{free}}}}, \label{MtimesNGlobe}
\end{equation}
where  ${\bf L}_N^{\rm{{\rm{free}}}}$ and ${\bf L}_M^{^{\rm{DD}}}$ can be thought of as the Laplacians of 1D
lattices with free and Dirichlet-Dirichlet  boundary conditions respectively.

The eigenvalues and eigenvectors of ${\bf L}_N^{\rm {\rm{free}}}$ and ${\bf L}_M^{^{\rm{DD}}}$ are well known \cite{IK},
\begin{align}
\Lambda_k &\equiv \Lambda_{m,n}=2r^{-1}(1-\cos{\theta_n})+2s^{-1}(1-\cos{2\varphi_m}),\label{LambdamnH}\\
\psi_{k,i} &\equiv \psi_{m,n}(x,y) = u_n(x)v_m(y), \\
\intertext{where $u_n(x),n=0,1,\dots,N-1$ are the eigenvectors of ${\bf L}_N^{\rm free}$,}
u_0(x)&=\sqrt{\frac 1 N}\quad\hbox{otherwise}\quad u_n(x)=\sqrt{\frac 2 N}\cos\left((x-\frac 12)\theta_n \right) \label{uvec} \\
\intertext{ and where $v_m(y),m=0,1,\dots,M-1$ are the eigenvectors of ${\bf L}_M^{^{\rm{DD}}}$,}
&v_m(y)=\sqrt{\frac{2}{M+1}}\sin(2 y \varphi_m)\label{varphikiH}
\end{align}
Here $\theta_n = \frac{\pi n}N$ and $\varphi_m=\frac{\pi(m+1)}{2M+2}$.

 The eigenvectors are orthogonal so
\begin{equation}
\sum_{x=1}^N u_{n_1}(x)u_{n_2}(x) =\d_{n_1,n_2},
\end{equation}
with similar formulae for $v_m(y)$. The inverse Laplacian \eqref{Lexpand} may now be written
\begin{equation}\label{invL}
{\cal L}_{i,j}^{-1} \equiv {\cL}_{x,y:x',y'}^{-1}= \sum_{m=0}^{M-1}\sum_{n=0}^{N-1} \frac{u_n(x)u_n(x')v_m(y) v_m(y')}{\Lambda_{m,n}}.
\end{equation}

\subsubsection{Evaluating the sums $\Sigma_1$ and $\Sigma_2$}
Let us now calculate the two   sums in \eqref{Sigmas}.
We need
\begin{equation}\label{usum}
\sum_{x=1}^N u_n(x) = \sqrt{N}\sum_{x=1}^N u_n(x) u_0(x) = \sqrt{N}\delta_{n,0}.
\end{equation}
 Noting that the sum $i=1$ to $N$ is equivalent to the sum $x=1$ to $N$ with $y=1$ we start by
 evaluating
\begin{align}
S(x',y')&\equiv \sum_{i=1}^N {\cL}_{i,j}^{-1}=\sum_{x=1}^M {\cL}_{x,1:x',y'}^{-1}\\
&=\sum_{x=1}^M \sum_{m=0}^{M-1}\sum_{n=0}^{N-1} \frac{u_n(x)v_m(1)u_n(x')v_m(y')}{\Lambda_{m,n}}=
\sum_{m=0}^{M-1}\frac{v_m(1)v_m(y')}{\Lambda_{m,0}}\\
&=\frac s{M+1} \sum_{m=0}^{M-1}\frac{\sin(2\varphi_m)\sin(2y'\varphi_m)}{1-\cos(2\varphi_m)}\\
&=\frac s{M+1} \sum_{m=0}^{M-1} \cot \varphi_m \sin (2y'\varphi_m) = \frac{(M+1-y')s}{M+1}.
\end{align}
The last equality is valid for all integer value of $y'$ in the range $y' \le 2M+1$, which is clearly our case.
The two required sums now follow:
\begin{align}
\Sigma_1 &= \sum_{j=1}^N\left(\sum_{i=1}^N {\cL}_{i,j}^{-1}\right)=\sum_{x'=1}^N S(x',1) =\frac {NMs}{M+1},\\
\Sigma_2&=S(x_1,y_1)-S(x_2,y_2) = \frac{(y_2-y_1)s}{M+1},
\end{align}
where in \eqref{Sigmas} the input node $\a =(x_1,y_1)$  and  the output node $\b =(x_2,y_2)$ and we have used \eqref{usum}.

 Substituting $\Sigma_1$ and $\Sigma_2$ into equation \eqref{resistorab1} the required resistance now takes the form
\begin{equation}\label{Rend}
R_{\alpha,\beta}= R_{x_1,y_1}^{x_2,y_2} =\frac{s(y_2-y_1)^2}{N(M+1)} +\hat R_{x_1,y_1}^{x_2,y_2}
\end{equation}

 \subsubsection{Evaluating $\hat R_{x_1,y_1}^{x_2,y_2}$}
We must first evaluate ${\cL}_{x,y:x',y'}^{-1}$ which is given by \eqref{invL}. If we were to substitute the eigenvalues and eigenvectors from \eqref{uvec}, \eqref{varphikiH} and \eqref{LambdamnH} the term $n=0$ would need to be treated separately. However notice that $u_N(x)=0$ and for $n=1$ to $N-1$, $u_{2N-n}(x)=-u_n(x)$ and $\Lambda_{m,2N-n }=\Lambda_{m,n}$ so
\begin{equation}
\sum_{n=1}^{N-1} \frac {u_n(x) u_n(x')}{\Lambda_{m,n}}=\sum_{n=N+1}^{2N-1} \frac {u_n(x) u_n(x')}{\Lambda_{m,n}}=\frac12\sum_{n=1}^{2N-1} \frac {u_n(x) u_n(x')}{\Lambda_{m,n}}.
\end{equation}
Now for $n=1$ to $2N-1$ let $w_n(x) =u_n(x)/\sqrt{2}$ and $w_0(x) =u_0(x)$ so
\begin{equation}
w_n(x) = \sqrt{\frac 1 N}\cos \left(\frac 12(2x-1)\theta_n\right ) \qquad \hbox{for}\qquad n=0,1\dots 2N-1.
\end{equation}
Equation \eqref{invL} now becomes
\begin{equation}\label{newinvL}
{\cL}_{x,y:x',y'}^{-1}= \sum_{m=0}^{M-1}\sum_{n=0}^{2N-1} \frac{w_n(x)w_n(x')v_m(y) v_m(y')}{\Lambda_{m,n}},
\end{equation}
and the $n=0$ term is no longer special.
Now, for integer $\ell$, we have the identity (\cite{W} equation (62))
\begin{equation}
\frac 1 {2N} \sum_{n=0}^{2N-1} \frac {\cos ( \ell \theta_n)}
 {{\rm ch}( 2 \Omega_m) - \cos \theta_n   } = \, \frac{{\rm ch}[2(N - \ell)\Omega_m)]}
{{\rm sh} (2\Omega_m ){\rm sh}(2 N \Omega_m)} ,\label{sumidentity}
\end{equation}
and
\begin{equation}
\Lambda_{m,n} = 2r^{-1}[ {\rm ch}(2\Omega_m) -\cos \theta_n],
\end{equation}
where we have introduced $ \rm {ch}(2\Omega_m)$ by
\begin{equation}\label{Lamch}
1 + h (1-\cos 2\varphi_m ) = {\rm ch} (2
\Omega_m)\qquad\hbox{or}\qquad {\rm sh}( \Omega_m) = \sqrt h \sin
\varphi_m.
\end{equation}
In order to evaluate ${\cL}_{x,y:x',y'}^{-1}$ we need
\begin{align}
\sum_{n=0}^{2N-1}\frac{w_n(x)w_n(x')}{\Lambda_{m,n}}
&= \frac 1 N \sum_{n=0}^{2N-1}\frac{\cos(\frac 12(2x-1) \theta_n)\cos(\frac12 (2x'-1)\theta_n)}{\Lambda_{m,n}}\\
&= \frac r {4N} \sum_{n=0}^{2N-1}\frac{\cos((x+x'-1)\theta_n)+\cos((x'-x)\theta_n)}{{\rm ch}( 2 \Omega_m) - \cos \theta_n  }\\
&=   r \frac{{\rm ch }(2(N-x-x'+1)\Omega_m)+{\rm ch }(2(N-x'+x)\Omega_m)}{2\,{\rm sh}( 2 \Omega_m) {\rm sh}( 2 N\Omega_m) },\\
\intertext{where we have used \eqref{sumidentity}, and combining
with \eqref{newinvL}} {\cL}_{x,y:x',y'}^{-1}&=   r
\sum_{m=0}^{M-1}v_m(y)v_m(y')\frac{{\rm ch
}((2N-2x'+1)\Omega_m){\rm ch }((1-2x)\Omega_m)}{{\rm sh}(
2\Omega_m) {\rm sh}( 2 N\Omega_m) }\label{Lfinal}
\end{align}
where we have assumed $x'\ge x$. Notice that the identity \eqref{sumidentity} has enabled the double sum \eqref{newinvL} to be reduced to a single sum.

From \eqref{Rhat}
\begin{equation}\label{Rhatfin}
\hat R_{\alpha,\beta}={\cal L}^{-1}_{x_1,y_1;x_1,y_1}   - {\cal
L}^{-1}_{x_1,y_1;x_2,y_2}- {\cal L}^{-1}_{x_2,y_2;x_1,y_1}+ {\cal
L}^{-1}_{x_2,y_2;x_2,y_2}
\end{equation}
Comparing the definitions \eqref{Lch} and \eqref{Lamch} we see
that $\Omega_m = L_{m+2}$. Notice also that
 $v_m(y)  = \sqrt{ \frac 2 {M+1}} S_{m+2}(y)$  .  Assuming $x_2\ge x_1$ and combining \eqref{Rend}, \eqref{Lfinal} and \eqref{Rhatfin} gives \eqref{Rhammockasym}.

\subsection{Method B: Using the recursion-transform technique of Tan  \cite{T11}}\label{sec:B:ham}

\begin{figure}[t]
\begin{center}
\includegraphics[scale=1.0]{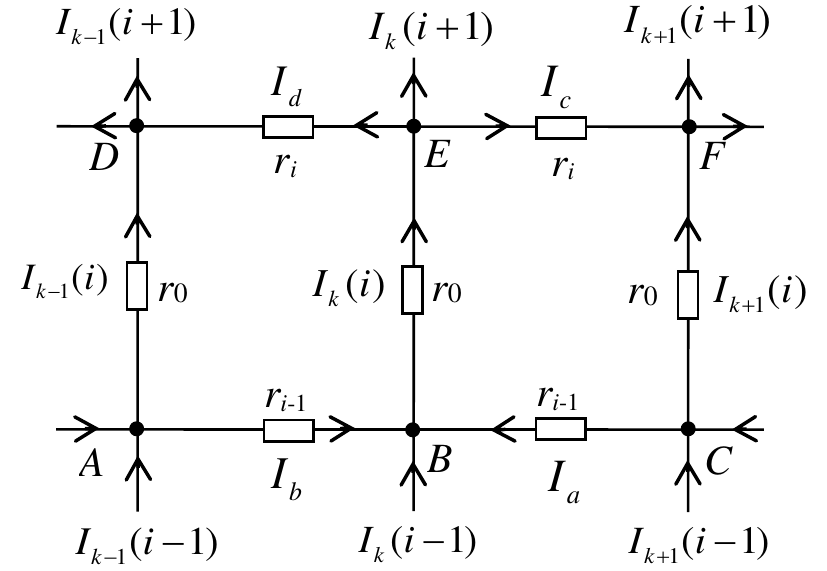}
\caption{\it The voltage loop ABEFCBEDA}   \label{loop}
\end{center}
\end{figure}

   We use the $k,s,t,p,q$ notation defined in section \ref{results}. This choice enables the use of symmetry and produces more symmetric coefficients \eqref{pqlab}.  

  Suppose that current $J$ is input at $N_p$ and flows out at $N_q$. Let $I_k(i)$ be the resulting radial current in the $i^{th}$ resistor from the lower edge of column $k$ in the direction of increasing $i$ (see figure \ref{loop}). Using Ohm's law  the potential difference between $N_p$ and $N_q$ may be measured along a path from $N_q$ to the common node $i=M+1$ and then to $N_p$ with the result
\begin{equation}\label{Res}
R_{x_1,y_1}^{x_2,y_2}= \frac{s} {J}\left( \sum _{i=y_2+1}^{M+1} I_q(i) -\sum _{i=y_1+1}^{M+1} I_{-p}(i)\right )
\end{equation}

\subsubsection{Relating the current distribution in three adjacent radial lines}

To determine the radial currents consider the voltage loop $ABEFCBEDA$, shown in figure \ref {loop}, centered on the $i^{th}$ resistor of the $k^{th}$ radial line .  If current $J$ enters at the node of height $y$ on the radial line $k=z$ charge conservation gives
\begin{align}
I_a+I_b &=I_k(i) -I_k(i-1) -J \d_{i,y+1}\d_{k,z} \label{Icon}\\
\hbox{and}\qquad  I_c+I_d&= I_k(i)-I_k(i+1) +J\d_{i,y}\d_{k,z}.
\end{align}
Here $z=q\, \hbox{or}-p$ and $y=y_1\, \hbox{or}\, y_2$. When $i=1$ in \eqref{Icon}, $I_k(i-1) = 0$.
The sum of the voltage differences round the loop is zero so using Ohm's law
\begin{equation}
s(2I_k(i) - I_{k-1}(i) -I_{k+1}(i)) +r_{i-1}(I_a+I_b) +r_i(I_c+I_d)=0,
\end{equation}
where $r_i=r$ for $1\le i\le M$, $r_0=r_{M+1}=0$.  
Combining these equations
\begin{align}
I_{k+1}(i) =&-h_{i-1} I_k(i-1)+ (h_{i}+h_{i-1}+2)I_k(i)   -h_{i} I_k(i+1) -I_{k-1}(i)\nonumber\\
 &+J(h_{i}\d_{i,y}-h_{i-1}\d_{i,y+1})\d_{k,z}
\label{Ieq}
\end{align}
where $h_i=r_i/s$. 
With $h=r/s$ equation \eqref{Ieq} may be written in matrix form
\begin{equation}\label{Ivec}
    I_{k+1}= [(2h+2)U_{M+1} - h W_{M+1}]I_k - I_{k-1} -hJ \d_{k,z}\epsilon_{i,y}
\end{equation}
where $U_m$ is an $m$-dimensional unit matrix, $\epsilon_y$ is a column matrix with $i^{th}$ element $\epsilon_{i,y}= \d_{i,y+1} -\d_{i,y}$ and
\begin{equation}\label{Wm}
{ W_{M+1} = \left (\begin{array}{lllllllll}
1&1&0&0&\dots&0&0&0\\
1&0&1&0&\dots &0&0&0\\
0&1&0&1&\dots &0&0&0\\
\vdots&\vdots&\vdots &\vdots&\dots&\vdots &\vdots&\vdots&\\
0&0&0&0&\dots&0&1&0\\
0&0&0&0&\dots& 1&0&1\\
0&0&0&0&\dots&0&1&1
\end{array} \right )}.
\end{equation}

For $k=t$ we only use the loop ABEDA in figure \ref{loop} to obtain the boundary equations
\hspace{-1.0in}
\begin{align}
I_{t-1}(i) &=(h_{i-1}+h_i +1)I_t(i)-h_{i-1}I_t(i-1)-h_i I_t(i+1) \\
\intertext{or in matrix form}
I_{t-1}&=[(2h+1)U_{M+1} -hW_{M+1}]I_t\label{Bt}
\end{align}
with a similar equation for $k=-s$.

 \subsubsection{The recurrence relation}

Let $\chi_i \equiv \varphi_{i-2} = \frac {(i-1)\pi}{2M+2}$. $W_{M+1}$ has eigenvalues $w_i = 2\cos(2\chi_i)$ and eigenvectors $\psi_i, i=1,2,\dots,M+1$. The $j^{th}$ component of $\psi_i$ is given by \cite{W}
\begin{equation}
\psi_i(j) = \cos (2j-1)\chi_i.
\end{equation}
 Let $\Psi$ be the matrix with $i^{th}$ row $\psi_i$ and define $X_k=\Psi I_k$. $\Psi$ is invertible with general element of the inverse
\begin{equation}
(\Psi^{-1})_{ij}=  \left \{
\begin{array}{lll}
\frac 1{M+1} &\hbox{for}&j=1\\
\qquad\\
\frac 2 {M+1} \cos (2i-1)\chi_j&\hbox{for}&2\le j \le M+1
\end{array}
\right ..
\end{equation}
 Using \eqref{Res}
\begin{equation}\label{RhammockX}
R_{x_1,y_1}^{x_2,y_2}= \frac{s}J\left(\sum_{i=1}^{M+1}  X_q(i) s_i(y_2) - \sum_{i=1}^{M+1}  X_{-p}(i) s_i(y_1) \right),
\end{equation}
where for $i>1$
\begin{equation}\label{si}
s_i(y) = \sum_{j=y+1}^{M+1}(\Psi^{-1})_{ji} = \frac 2 {M+1} \sum_{j=y+1}^{M+1} \cos (2j-1)\chi_i = -\frac 1{M+1} \frac{\sin(2y\chi_i)}{\sin\chi_i}
\end{equation}
and $s_1(y)= (M+1-y)/(M+1)$.

Multiplying \eqref{Ivec}
on the left by $\Psi$, noting that $\Psi W_{M+1}$ is diagonal with diagonal elememts $w_i$, and taking the $i^{th}$ component
\begin{equation}\label{Xeq}
X_{k+1}(i) = u_i X_k(i) -X_{k-1}(i) -hJ\d_{k,z}\zeta_i(y)
\end{equation}
where $u_i = 2h+2 -h w_i$ and
\begin{equation}
\zeta_i(y)\equiv \psi_i (y+1)-\psi_i(y)= -2 \sin(2y\chi_i) \sin \chi_i.
\end{equation}
 Applying $\Psi$ to \eqref{Bt} and taking the $i^{th}$ component
\begin{equation}
 X_{t-1}(i) = (u_i-1)X_t(i) \qquad \hbox{and similarly} \qquad  X_{-s+1}(i) = (u_i-1)X_{-s}(i).
\label{sandt}
\end{equation}

\subsubsection{Solving the recurrence relation}

For $k\ne z$ the general solution of \eqref{Xeq} is a linear combination of $\l_i^k$ and $\bar \l_i^k$ where
where $\l_i$ and $\bar \l_i$ are solutions of \eqref{laeq} in terms of which $\l_i +\bar \l _i = u_i$ and $\l_i \bar \l_i=1$.
The  coefficients depend on the region.
\begin{align}\label{Xsol}
X_k(i) &= A_i \l_i^k + \bar A_i \bar \l_i ^k \qquad \hbox{for} \qquad -p\le k \le q\\
X_k(i) &= B_i \l_i^k + \bar B_i \bar \l_i ^k \qquad \hbox{for} \qquad q\le k \le t \label{XsolB}\\
X_k(i) &= S_i \l_i^k + \bar S_i \bar \l_i ^k \qquad \hbox{for} \qquad -s\le k \le -p\label{XsolC}.
\end{align}
Matching the solutions at $k=q$ and $k=-p$,
\begin{align}
(A_i-B_i)\l_i^{2q} +\bar A_i- \bar B_i&=0\qquad \hbox{and} \qquad (\bar A_i -\bar S_i) \l_i^{2p} +A_i-S_i=0.\label{sim1}
\end{align}
Subsituting in the boundary equations \eqref{sandt}
\begin{equation}
\bar B_i =B_i \l_i^{2t+1} \qquad \hbox{and} \qquad S_i = \bar S_i \l_i^{2s+1}.\label{sim2}
\end{equation}
The final two relations arise from the $k=q$ and $k=-p$ radial lines where the current $J$ is input and output. Using \eqref{Xeq} with $k=z=q$ and secondly with $k=z=-p$, in the second case replacing $J$ by $-J$
\begin{equation}
\bar B_i- \bar A_i = -\frac{ hJ \l_i^q \zeta_i (y_2)}{\l_i- \bar \l_i} \qquad \hbox{and} \qquad \bar S_i -\bar A_i= -\frac{ hJ \bar \l_i^p \zeta_i (y_1)}{\l_i- \bar \l_i}.
\label{sim3}
\end{equation}
Solving equations \eqref{sim1}, \eqref{sim2} and \eqref{sim3} for $A_i$ and $\bar A_i$ and substituting in \eqref{Xsol} gives for $1<i \le M+1$
\begin{equation}
X_q(i) = \frac {hJ [\a \zeta_i(y_2)-\b \zeta_i(y_1)]}{(\l_i-\bar \l_i)D_i}\quad \hbox{and}\quad
X_{-p}(i) = -\frac {hJ [\g \zeta_i(y_1)-\b\zeta_i(y_2)]}{(\l_i-\bar \l_i)D_i}\label{X2}
\end{equation}
where $D_i=\l_i^{ n} - \l_i^{-n}$.
\begin{align}\a&=(\l_i^{t-q+\ha} + \bar \l_i^{t-q+\ha})(\l_i^{s+q+\ha} + \bar \l_i^{s+q+\ha})\\
\b&= (\l_i^{t-q+\ha} + \bar \l_i^{t-q+\ha})(\l_i^{s-p+\ha}+\bar \l_i^{s-p+\ha})\\
\g&= (\l_i^{t+p+\ha}+\bar \l_i^{t+p+\ha})(\l_i^{s-p+\ha}+\bar \l_i^{s-p+\ha}).
\end{align}
Now $\chi_1=0$ and $\l_1=1$ so the above expressions are indeterminate when $i=1$, but this may be resolved by taking limits.
\begin{equation}\label{Xlim}
X_q(1) = X_{-p}(1) = -\frac{J}{N}(y_2-y_1).
\end{equation}
Substituting \eqref{X2} and \eqref{Xlim} into \eqref{RhammockX} gives the required result \eqref{Rhammockasym}

\section{Summary and Discussion}
\label{end}

We have derived the resistance between two arbitrary nodes of the ``hammock'' network using the two different methods, A and B.

Instead of focussing on the potentials as in the Laplacian approach of Method~A, the recursive strategy in Method~B is to obtain a relation between the vertical currents in three adjacent columns.

Besides different starting strategies, we use different coordinate notations for the different approaches. The co-ordinates used in method B enable the use of symmetry and  lead to   symmetric coefficients \eqref{pqlab}.

Each approach has its advantages and disadvantages in general. 
For {\emph{Method A}}, the formula \eqref{Rab} for the two-point resistance is valid for an arbitrary network. 
The two-point resistance can be computed for cubic lattices in any spatial dimension (since the Laplacian for $d$-dimensional regular square lattices can be represented as the sum of $d$ one-dimensional Laplacians, with known eigenvalues and eigenvectors) under various boundary conditions \cite{W}, for example free or periodic. 
Thus the resistance problem is one of the few non-trivial problems which can be solved exactly in high dimensions. 
Once the eigenvalues and eigenvectors of the Laplacian are known, the resistance between two arbitrary points is given by a very simple summation formula \eqref{Rab}. 
While the determination of the eigensystem is straightforward to obtain for hypercubic lattices in any dimensions, the approach cannot readily deal with other complex graphs.
However for the square lattice with one or two added nodes the Laplacian may be replaced by its the first or second minors respectively, for example, as in Method A.

In terms of applications to the ``hammock'' network, conversion to a rectangular network is an essential part of both methods. 
In method~A this is so that the decomposition \eqref{decomp} may be used. In method B , the ``hammock'' is  extended to a full rectangle with $N$ columns and $M+2$  rows with zero resistance in the top and bottom rows. If $r_i$ is the value of the resistors in row $i$ of the rectangle then $r_0=r_{M+1}=0$ and otherwise $r_i=r$ so that the same recursion \eqref{Ieq} can be used for all rows.  This extension is not possible in method~A since the coefficients are conductances and would be infinite in the top and  bottom rows. Instead the contribution of the two additional nodes is first separated to yield \eqref{resistorab}. 
 
Both methods use the eigenvectors and eigenvalues of the Laplacian  ${\bf L}^{{\rm{free}}}_m$ of the linear chain of length $m$ with free boundaries. Method~A further requires the eigensystem for the Laplacian of a chain with Dirichlet-Dirichelet boundary conditions. This leads to a double sum \eqref{newinvL} and in order to arrive at the final formula one of the sums has to be  removed using a non-trivial identity \eqref{sumidentity}. Reference is made to Wu \cite{W} for the proof of the identity. The summation which occurs in the final formula is the starting point of method B  \eqref{Res} and the summand involves the transformed current vector and the inverse of the eigenvector matrix (see \eqref{RhammockX}). The former
requires the solution of a recurrence relation with constant coefficients and the latter involves a trivial summation \eqref{si}.
Finally, method~A requires reference to previous calculations \eqref{resistorab},\eqref{sumidentity} whereas method~B is virtually self contained using only Ohm's law and Kirchhoff's laws.


The Laplacian approach of {\emph{Method A}} has so far delivered analytical formulae for the two-point resistances for classes of graphs such as 
regular two-dimensional square lattices under different boundary conditions \cite{W}; 
higher dimensional regular square lattices \cite{W}; regular square lattices with a single  additional node;
so called cobweb \cite{IKW} and fan \cite{IKfan} networks; 
and regular square lattices with two additional nodes -- the so-called globe network \cite{IK}.

{\emph{Method B}}  has previously been applied to the  fan  \cite{ETW}, cobweb \cite{ETW2}  and globe networks \cite{TEW}. The method  has also been used on the regular square lattice but the potential difference along the top edge is non-zero and has to be calculated by interchanging the $x$ and $y$ axes; alternatively the required potential difference may determined along a vertical path followed by a horizontal path \cite{JWE}. 

Method B could also be applied to problems where the horizontal resistance depends in different ways on the row index $i$. The simplest of these would be $r_i= i r$ which would apply to a fan network embedded in the plane where the length of the resistor wires would be proportional to the distance from the apex. The method as presented here would then require finding the eigensystem of a tridiagonal matrix with elements depending on $r_i$. This can be avoided by working with the vector $I(i)\equiv \{I_1(i),I_2(i),\dots I_N(i)\}$ which when transformed would lead to a second order recurrence relation with coefficients depending on~$i$.
 
\vspace{1cm}

{\bf{Acknowledgement:}} This work were supported by a Marie Curie IIF (Project no. 300206-RAVEN)and IRSES (Projects no. 295302-SPIDER and 612707-DIONICOS) within 7th European Community Framework Programme and by the grant of the Science Committee of the Ministry of Science and Education of the Republic of Armenia under contract 13-1C080.

\vspace{1cm}

{\bf \large References.}

\begin{enumerate}

\item\label{disorder}
Kirkpatrick S., Rev. Mod. Phys., 1973, 45, 574.
Derrida B., Vannimenus J., J. Phys. A, 1982, 15, L557.
Harris A.B., Lubensky T.C., Phys. Rev. B, 1987, 35, 6964.

\item\label{DS} Doyle P G and Snell J L, "Random Walks and Electrical Networks", The Mathematical Association of America, Washington, DC, 1984.

\item\label{Redner}
Redner S., A Guide to First-Passage Processes, Cambridge University Press, Cambridge, 2001.

\item\label{Kat}
Katsura S., Morita T., Inawashiro S., Horiguchi T., Abe Y., J. Math. Phys., 1971, 12, 892.

\item\label{KR} Klein D J and Randić M, "Resistance distance", J. Math. Chem. {\bf 12} 81–95 (1993).

\item \label{K} Kirchhoff G, Ann. Phys. Chem. {\bf 72} 497-508 (1847).

\item\label{XG} Xiao W J and Gutman I, "Resistance distance and Laplacian spectrum", Theory Chem. Acc. {\bf 110}
284–9 (2003).

\item\label{W} Wu F Y, "Theory of resistor networks: the two-point resistance", J. Phys. A: Math. Gen. {bf 37} 6653 (2004).

\item\label{IKW} Izmailian N Sh, Kenna R and Wu F Y, "The two point resistance of a resistor network: A new formulation and application to the cobweb network", J Phys A:Math.Theor. {\bf 47} 035003 (10pp) (2014).

\item\label{IK} Izmailian N Sh and Kenna R, "Generalised formulation of the Laplacian approach to
resistor networks", J. Stat. Mech. (2014) P09016; arXiv:1406.5045
[math-ph].

\item\label{SR} Seshu S and Reed M B, "Linear Graphs and Electrical Networks", Addison-Wesley, Reading, Mass, 1961.

\item\label{SS} Sharpe G E and Styan G P H, "Circuit duality and the general network inverse", IEEE Trans. Circuit Theory {\bf 12} 22–27 (1965).

\item\label{BGX} Bapat R B, Gutman I and Xiao W J, "A simple method for computing resistance distance",
Z. Naturforsch. {\bf 58a} 494–8 (2003).

\item\label{T11} Tan Z-Z, "Resistance Network Model", (Xi'an : Xidian University Press) pp9-216 (2011).

\item\label{TZY} Tan Z-Z, Zhou L and Yang J-H,  "The equivalent resistance of $3 \times n$ cobweb network and its
conjecture of an $m \times n$ cobweb network", J Phys A: Math. Theor. {\bf 46} 195202 (12pp) (2013).

\item\label{TEW} Tan Z-Z, Essam J W and Wu F Y, "Two-point resistance of a resistor network embedded
on a globe", Phys. Rev. E {\bf 90} 012130 (2014).

\item\label{YK} Yang Y J and Klein D J, "A recursion formula for resistance distances and its
applications", Disc. Appl. Math. {\bf 161} 2702 (2013).

\item\label{IKfan} Izmailian N Sh and Kenna R, "The two-point resistance of fan networks", arXiv:1401.4463
[cond-mat].

\item\label{ETW}  Essam J W, Tan Z-Z and Wu F Y, "Proof and extension of the resistance formula for an $m \times n$ cobweb network conjectured by Tan, Zhou and Yang", arXiv:1312.6727 [cond-mat].

\item\label{ETW2} Essam J W, Tan Z-Z and Wu F Y, "Resistance between two nodes in general position on an $m\times n$ fan network", Phys. Rev. E, {\bf 90} 032130 (2014), arXiv:1404.2828 [cond-mat].




\item\label{JWE} Essam J W, unpublished.

\end{enumerate}
\end{document}